# SPEAR: Structure–Property Explainability with Attention Regularization

*Aditya Raghavan[1], Utkarsh Pratiush[1], Dalton A. Pearl[1], Jade Holliman Jr[1], Katharine Page[1], Philip D Rack[1], Sergei V Kalinin[1]*

[1]University of Tennessee, Knoxville, TN, USA

araghav4@utk.edu, sergei2@utk.edu

**Abstract**

Machine learning models are increasingly used to learn structure–property relationships from spectroscopic and diffraction data, yet their adoption in materials discovery is often limited by poor interpretability of model predictions. Although attention mechanisms are frequently treated as inherently explainable, unregularized attention in practice can yield unstable, fragmented, or intensity-driven attribution patterns that obscure the physical origin of structure–property relationships. Here, we introduce SPEAR (Structure–Property Explainability with Attention Regularization), a framework that explicitly constrains attention distributions during training to improve their stability, selectivity, and physical interpretability. SPEAR augments attention-based regression with a learnable temperature that controls attention concentration and a smoothness penalty that enforces coherence across neighboring spectral positions, treating attention as a learnable explanatory object rather than a post hoc visualization. Using synthetic spectral benchmarks with known generative structure, we show that attention regularization produces smooth, contiguous attribution profiles aligned with causal features while preserving predictive accuracy. Applied to experimental X-ray diffraction data from a combinatorial rare-earth zirconate thin-film library, the regularized attention model exhibits target-aware attribution that selectively emphasizes physically relevant diffraction features and decouples feature importance from raw peak intensity. These results demonstrate that attention regularization provides a principled training constraint for explainable structure–property regression, yielding stable and mechanistically meaningful explanations without sacrificing predictive performance. The reflection identified by the regularized model prompted a reassessment of our earlier structural analysis, revealing a correlation between the 220 peak position, the tetragonal distortion that accommodates cation size disorder, and the local thermal conductivity.

## I. Introduction

A central goal of modern materials science is to understand why a material behaves the way it does — why one composition behaves differently from another, why a particular atomic arrangement is more stable, and why shifting a synthesis parameter changes a property. Machine learning (ML) has become a powerful engine for **predicting** these behaviors from experimental data: given a spectrum, diffraction pattern, synthesis parameters or even a microscope image, a trained model can often predict a target property with high accuracy.[1-4] Yet accurate prediction is not the same as physical understanding. When a model is asked not just to predict but to **explain** which structural quantities influenced its answer, the explanation is often unreliable, pointing to the wrong part of the data or tracking irrelevant signal intensity rather than peaks associated with the answer.[5-7] If a researcher trusts such an explanation, they may pursue the wrong composition, synthesis condition or material descriptor, wasting significant experimental resources or arriving at incorrect physical conclusions.[8, 9]

The difficulty is sharpest for spectroscopic and diffraction data. In an X-ray diffraction (XRD) pattern, for example, a material's atomic structure encodes itself as a set of peaks: their positions encode lattice spacing, their widths encode disorder, and their relative intensities depend on the arrangement of atoms within the unit cell.[10-12] When composition varies across a combinatorial library of samples, many peaks can shift together, because they are physically correlated, which makes it difficult for a standard attribution method to isolate which peak actually drives a given property change, as opposed to which peak happens to co-vary with it.[13, 14] In high-throughput or autonomous discovery workflows, model explanations are not just visualizations: they are scientific hypotheses that guide the next experiment, the next synthesis target, and the next characterization decision.[15-18] A diffuse or misleading attribution does not merely fail to inform; it actively misdirects the subsequent theory development or decision making.[8, 19]

Machine learning has become a foundational component of contemporary materials discovery and characterization workflows. [1-3, 20, 21] Across length scales and materials classes, ML models now routinely learn relationships between experimental measurements such as spectra, diffraction patterns, microscope images, and composition data, and target material properties such as conductivity, and mechanical performance.[22-29] More broadly, these models enable closed-loop workflows where model predictions guide the next synthesis or measurement step, accelerating screening and optimization, and support the identification of new compositions likely to meet a target specification.[15-18, 30-33]

As ML workflows are deployed in increasingly consequential scientific and technological settings, explainability has emerged as a central requirement rather than a secondary convenience.[5-7, 34-38] A model that predicts accurately but cannot explain its predictions has limited scientific value: it may be used for screening, but it provides limited insight into which spectral regions are physically responsible for the predicted property.[5, 6, 9] In materials science, where models are routinely used to propose new compounds, microstructures, and processing conditions, the absence of reliable interpretability does not merely limit adoption, it also constrains what can be learned from the data.[5, 6, 8, 9, 26] Understanding why a model makes a particular prediction is therefore critical for trust and validation,[7, 8] for extracting physical insight,[5, 6] and for directing subsequent experiments.[8, 39, 40]

Established methods for quantitative XRD analysis such as Rietveld refinement, whole-pattern fitting, and reference intensity ratio approaches are highly effective for structural characterization when the phases present are known and the number of samples is small.[41, 42] In these settings, Rietveld refinement in particular extracts rich structural information: lattice parameters, phase fractions, atomic positions, and microstructural quantities, all with well-understood uncertainty.[41] However, for compositionally complex disordered systems such as the disordered rare-earth zirconates studied here, identifying a valid starting structural model is itself non-trivial. The challenge arises in combinatorial and high-throughput studies, where hundreds or thousands of diffraction patterns are collected across continuous composition gradients and the phase space may be only partially known. In this regime, manual refinement does not scale and the goal shifts from detailed structural determination of individual samples to extracting composition–structure–property trends across an entire library.[43, 44] Machine learning methods are well suited to this regime: NMF-based approaches and convolutional neural networks can process large pattern sets rapidly and without requiring a pre-specified structural model, achieving high accuracy in phase identification and property prediction at scale.[8, 13, 14] What they do not yet provide is a reliable account of which spectral regions drove a given prediction. This interpretability gap, rather than any limitation in predictive accuracy, is what SPEAR is designed to address.[5, 6, 10, 11]

In current practice, explainable ML is most commonly operationalized through SHAP (SHapley Additive exPlanations), a family of post hoc attribution methods that assign each input feature a contribution to an individual prediction by estimating how much each spectral region contributes to the prediction.[45] SHAP provides additive, locally consistent explanations and is widely adopted for its model-agnostic variants and intuitive feature-importance visualizations, with demonstrated applications across chemistry and materials science.[45-52] However, diffraction and spectroscopic data are characterized by strongly correlated spectral regions, and in these settings SHAP attributions can be sensitive to the choice of background distribution and can diffuse importance across correlated spectral regions in physically ambiguous ways.[13, 14] In practice, this means that a SHAP analysis of a diffraction model may assign high importance to multiple peaks that co-vary with composition even when only one peak is physically responsible for the target property, making it difficult to identify the mechanistically relevant structural feature. Moreover, SHAP is applied separately, after the model has already been trained, and does not influence how the model learnt, meaning the resulting explanations are not guaranteed to be stable or physically consistent.[8, 53, 54]

When an experienced crystallographer interprets a diffraction pattern, they do not treat every point on the spectrum equally; they focus on reflections whose positions, profiles, and relative intensities encode the crystallographic information relevant to the structure under study. Peak positions constrain lattice parameters and symmetry, peak broadening provides insight into crystallite size and microstrain, and systematic intensity variations reflect preferred orientation, atomic site occupancies, and other structural characteristics. Attention mechanisms encode this same selectivity into a machine learning model. During training, the model learns to assign a weight to each point in the spectrum, reflecting how much it relies on that region when making a prediction.[55, 56] After training is complete, these weights can be plotted directly over the spectrum, producing a map that shows which reflections the model considered most important for the selected property. This is what makes attention attractive as an explanation tool: rather than requiring a separate post-hoc analysis, the model's focus is readable directly from its internal computation.[57, 58] Attention mechanisms are thus a natural candidate for explainable structure–property regression, and unlike post hoc attribution methods, they are an integral part of the predictive computation rather than an external wrapper applied after training. Because these weights are normalized and can become concentrated, they provide a direct map of which reflections or spectral regions the model relies on when making a prediction.[57-61]

Despite these advantages, a critical limitation of standard, unregularized attention is that it frequently produces spiky, noisy, and unstable attribution patterns that bear little relation to the underlying physics.[60, 62-64] Unlike post hoc methods applied after training, attention is part of the prediction itself, meaning the model's focus can be shaped and constrained during training to produce weights that are physically meaningful from the outset.[8, 19, 38, 65-67] In practice, however, plain attention tends to focus on whichever peaks are most intense in the spectrum rather than those most physically relevant to the target property, and this focus is often inconsistent, shifting unpredictably across samples that are structurally similar. [60, 62, 64] The key question is therefore not whether attention can provide interpretable explanations, but under what conditions it reliably does.

In this work, we introduce SPEAR (Structure–Property Explainability with Attention Regularization), a framework that moves beyond simply reading attention weights after training and instead actively shapes the attention distribution during learning, so that the resulting weights are spatially coherent and physically meaningful from the start. We formalize two complementary regularization mechanisms: (i) a parameter that controls how narrowly or broadly the model focuses its attention, preventing it from locking onto a single spectral channel or treating the entire spectrum as equally important;[67] and (ii) a smoothness penalty that enforces coherence across neighboring spectral positions, preventing attention from jumping erratically between adjacent points.[68] We evaluate SPEAR on synthetic spectral benchmarks constructed from superpositions of Gaussian peaks with known generative structure, which allows controlled, ground-truth assessment of attribution quality. We further validate the approach on experimental XRD data from a combinatorial $(Gd_xDy_yHo_zEr_{1-x-y-z})_2Zr_2O_7$ thin-film library, hereafter referred to as the $(Gd,Dy,Ho,Er)_2Zr_2O_7$ library, where the regularized attention model is tested on both a controlled feature-regression task (predicting diffraction peak position) and a property-prediction task with distributed structural contributions (predicting thermal conductivity). These results show that with the right constraints, a model's attention can be made to mean something — pointing reliably to the diffraction features that matter and providing a foundation for physically grounded hypothesis generation in materials discovery.

## II. Method Description:

Attention mechanisms were introduced to enable neural networks to compute input-dependent importance weights, allowing models to selectively aggregate relevant information rather than relying on fixed-length representations,[56] and were later popularized by the Transformer architecture based entirely on self-attention.[55] Attention produces instance-specific weight distributions over inputs, which has led to its widespread use as an interpretability signal, for example through visual attention maps in vision models and attention rollout methods in Transformers.[69, 70] While attention weights do not universally guarantee causal explanations, prior work shows they can provide meaningful, hypothesis-generating insight into model behavior when interpreted carefully or constrained appropriately.[57, 58, 60] Unlike causal inference approaches that explicitly reconstruct causal relationships[71, 72], attention-based attribution is interpreted here as predictive sensitivity rather than causal effect.

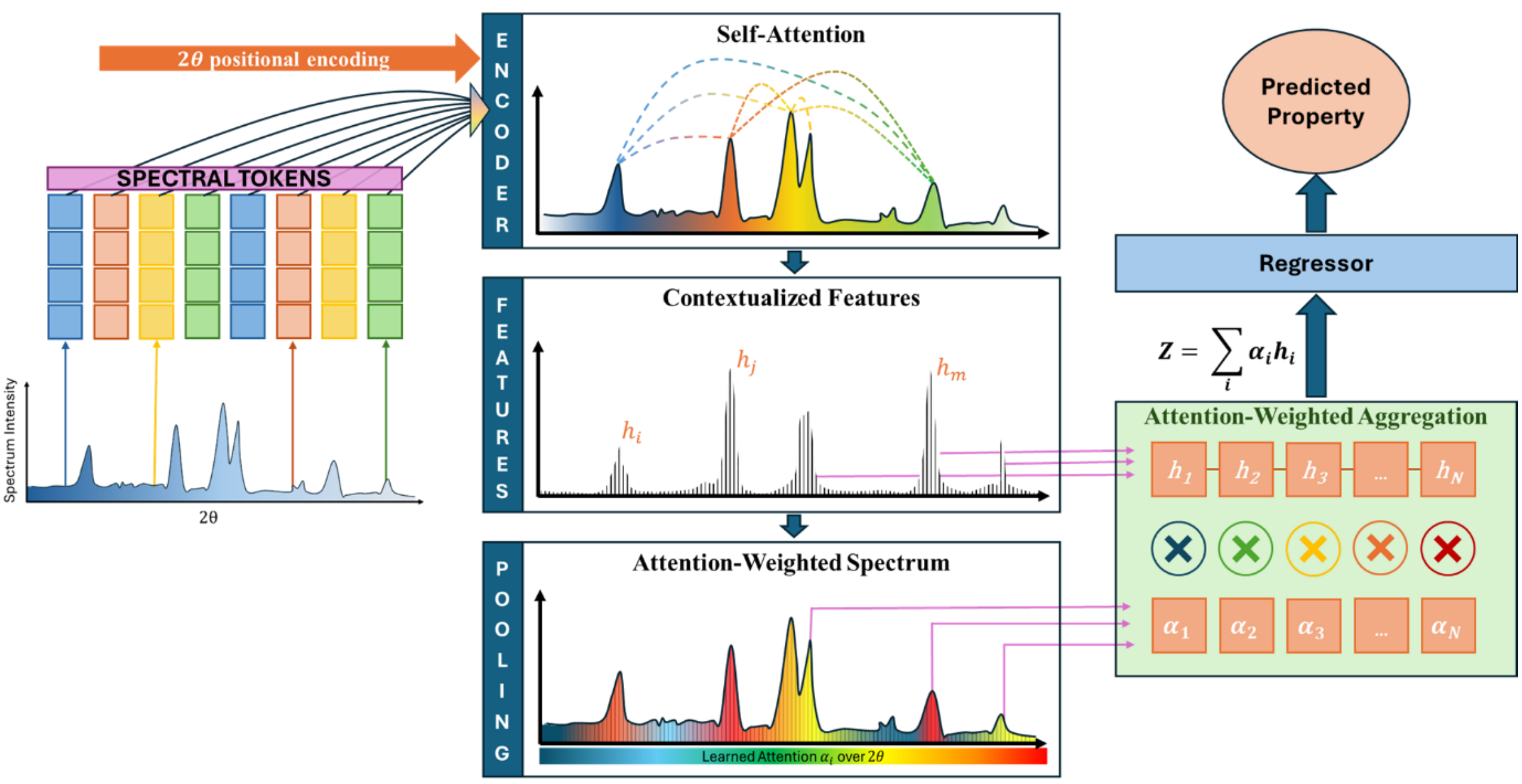

**Figure 1. Schematic of the workflow:** The figure illustrates the use of learned attention weights to aggregate encoded spectral features prior to regression, with the resulting attention distribution providing an interpretable weighting over spectral regions.

In this work, we adopt attention as a structured pooling mechanism for spectral regression, as schematically illustrated in **Fig. 1**. The input spectrum is first discretized into spectral tokens with positional encoding and mapped to latent feature representations by an encoder. Learned attention weights are then used to perform attention-weighted aggregation over these features prior to regression, allowing the model to selectively emphasize informative spectral regions while yielding an interpretable attention distribution that reflects their relative contribution to the predicted property.

**II A. Experimental Dataset and Preprocessing**

Synchrotron X-ray diffraction measurements were acquired across a combinatorial $(Gd,Dy,Ho,Er)_2Zr_2O_7$ thin-film library to characterize structural variations associated with rare-earth composition gradients (see Ref. 73 for synthesis, composition, and measurement details). The diffraction patterns exhibit broad features consistent with a disordered defect-fluorite structure, as commonly observed in compositionally complex rare-earth zirconates with an average rare-earth (RE) to zirconium ionic-radius ratio $r_{\mathrm{RE}}/r_{\mathrm{Zr}} < 1.46$, with observed reflections indexed as $\{111\}$, $\{200\}$, $\{220\}$ and $\{311\}$ in increasing order of $2\theta$. Systematic shifts in peak positions and changes in relative peak prominence are observed across the library, consistent with spatially varying lattice spacing and disorder-related structural variation. For the present analysis, each measured wafer location was treated as a discreet composition, and the processed one-dimensional XRD spectrum at that location was used as the model input.

The spectra were matched to the metadata file by measurement point ID, and the wafer coordinates were converted from millimeters to centimeters. The stacked XRD intensity matrix was globally normalized to the range [0,1], followed by per-spectrum mean subtraction to remove background-like offsets. Negative values after mean subtraction were clipped to zero. The spectra were then downsampled along the $2\theta$ axis by a factor of 15 and restricted to the diffraction-angle window between 2.22-8.92° for model training. Because the computational cost of self-attention scales quadratically with sequence length, this reduction is what makes training and attribution tractable on a full combinatorial library, and it is a prerequisite for the method being usable for on-the-fly analysis during data collection rather than only in post hoc studies. To reduce differences in dynamic range between spectra before transformer training, the retained intensity values were shifted to be non-negative, scaled by the 95th-percentile intensity of each spectrum, and transformed using $\log(1 + I)$ compression. Each spectral token was represented using two features: the processed intensity value and the corresponding normalized $2\theta$ coordinate. This representation preserves variations across the full diffraction profile rather than reducing each pattern to a small set of fitted descriptors. The center region of the wafer was excluded because of a known experimental artifact that changes the preferred orientation of the center compositions (see Ref. 73 for details).

Two scalar regression targets were considered. In the first task, the target was the $2\theta$ position of the {200} reflection, the second diffraction peak in the pattern, extracted from each measured spectrum by peak fitting using LMFit.[74] This target provides a controlled feature-regression setting because the relevant spectral feature is known by construction. The second peak was selected because it is not the most intense feature in the diffraction pattern, allowing us to test whether the model localizes attribution to the target-defining peak rather than assigning importance primarily according to peak magnitude. In the second task, experimentally measured thermal conductivity values across the composition space were used as the regression target. This property-regression task examines whether attention regularization produces stable and interpretable attribution patterns when the target is expected to depend on distributed structural information encoded across multiple diffraction features rather than on a single fitted peak position.

Thermal conductivity is expected to correlate with a combination of crystallographic features rather than a single diffraction peak. Features such as systematic peak shifts (lattice parameter changes), isotropic or anisotropic peak broadening (microstrain and defect density), diffuse scattering (local structural disorder), and subtle changes in relative reflection intensities associated with phase composition or symmetry changes / ordering may all serve as indirect indicators of structural characteristics that will influence phonon transport. An attention mechanism that can identify multiple regions or features of the diffraction pattern is therefore appropriate for the anticipated multivariate structural origins of thermal conductivity.

**II B. Problem Setup and Model Overview**

We consider a supervised regression problem in which a scalar target property is predicted from spectral inputs. The input spectra are the XRD patterns described in section IIa. For the transformer model, these ordered measurements are treated as sequence, preserving the physical continuity of neighboring diffraction features.

We train a transformer-based regressor $f_\theta$, parameterized by $\theta$, to map the input spectrum to a scalar output. In the synthetic benchmarks used for controlled validation, the target corresponds to the amplitude of a Gaussian peak embedded within the spectrum.

The transformer encoder produces contextualized representations through self-attention layers. To aggregate information across spectral positions for regression, we employ an attention-based pooling mechanism that assigns a normalized weight to each position and forms a weighted sum of the corresponding hidden states.

**II C. Plain vs. Learnable Attention Pooling**

In this work, we distinguish between *plain attention* and *SPEAR* within the attention-based pooling framework used for regression. By *plain attention*, we refer to the standard attention pooling mechanism commonly used in transformer regressors, in which attention weights are obtained via a *softmax* over attention logits and optimized solely through the prediction loss. In this setting, the attention distribution is unconstrained: its sharpness, locality, and stability emerge implicitly from optimization but are not explicitly controlled.

In contrast, the proposed SPEAR framework treats attention as a learnable explanatory object whose distributional properties are shaped during training. Rather than interpreting attention *post hoc*, SPEAR introduces explicit regularization terms that control both the concentration and smoothness of the attention weights. This formulation allows the model to adapt how selectively it attends to different input positions while enforcing physically motivated constraints, yielding attention distributions that are more stable and interpretable.

Importantly, both plain and regularized attention models share the same backbone architecture and differ only in the presence or absence of attention regularization terms, enabling a controlled comparison of interpretability without confounding architectural changes.

### II D. Attention Formulation

The attention weighting shown schematically in Fig. 1 is formalized through scaled dot-product attention. The transformer encoder computes contextualized representations using scaled dot-product self-attention. For a batch index $b$, attention head $h$, query position $i$, and key position $j$, the attention logits are defined as:

$$z_{b,h,i,j} = \frac{\langle \boldsymbol{q}_{b,h,i}, \boldsymbol{k}_{b,h,j} \rangle}{\sqrt{d_j}} \tag{1}$$

where $\mathbf{q}_{b,h,i}$ and $\mathbf{k}_{b,h,j}$ denote the query and key vectors associated with positions $i$ and $j$, respectively, and $d_k$ is the dimensionality of the key vectors used for scaling.

The attention weights are obtained by applying a softmax normalization over the key dimension:

$$\alpha_{b,h,i,j} = \frac{exp(z_{b,h,i,j})}{\sum_{j'=1}^{K} exp(z_{b,h,i,j'})} \tag{2}$$

For a fixed batch element $b$, head $h$, and query position $i$, the attention weights $\alpha_{b,h,i,j}$ form a probability distribution over key positions $j$. These weights quantify the relative importance of different spectral positions when constructing the contextual representation at position $i$.

`

In the absence of any additional constraints, this formulation corresponds to standard attention as used in transformer-based regression models. In the following section, we introduce regularization terms that explicitly shape the concentration and smoothness of this attention distribution to improve interpretability.

**II E. Loss Terms**

When trained using prediction loss alone, attention-based regression models can learn attention distributions that are unstable or poorly conditioned, such as becoming excessively sharp, overly uniform, or exhibiting high-frequency variations across neighboring inputs. While these behaviors may not strongly affect predictive accuracy, they undermine the interpretability of attention as a meaningful importance weighting.

To address this, we augment the standard prediction loss with explicit regularization terms that control the concentration and smoothness of the attention distribution, encouraging attention patterns that are both well-behaved and interpretable during training.

The first component of the attention-based regression model is the prediction loss term. Given model outputs $\hat{y}_i = f_\theta(x_i)$ for inputs $x_i$ and ground-truth labels $y_i$ , we use the mean squared error (MSE) loss:

$$\mathcal{L}_{\text{pred}}(\boldsymbol{\theta}) = \frac{1}{N}\sum_{i=1}^{N} (\hat{y}_i - y_i)^2 \quad (3)$$

This term ensures that the model accurately learns the target property (e.g., the Gaussian peak amplitude in the synthetic setting).

The second component that we introduce in this work is the attention concentration control. Here, to explicitly distinguish SPEAR from plain attention pooling, we introduce a learnable temperature parameter $\tau$ that rescales the attention logits prior to softmax normalization. In standard attention pooling, the softmax temperature is fixed, implicitly enforcing a single scale of selectivity. Here, $\tau$ is optimized jointly with model parameters, allowing the network to adaptively control how concentrated or diffuse its attention should be.

$$\alpha_{b,h,i,j}(\tau) = \frac{\exp(z_{b,h,i,j}/\tau)}{\sum_{j'=1}^{K}\exp(z_{b,h,i,j'}/\tau)} \tag{4}$$

Smaller values of $\tau$ promote more concentrated attention, while larger values yield more diffuse weighting. To prevent degenerate solutions in which attention becomes either excessively sharp or overly uniform, we impose a weak prior on $\tau$, implemented as a quadratic penalty on $\log\tau$ relative to a reference value $\tau_0$:

$$\mathcal{L}_{\tau}(\boldsymbol{\theta}) = \lambda_{\tau}(\log\tau - \log\tau_0)^2 \tag{5}$$

This term encourages adaptive but bounded control over attention concentration.

Finally, to discourage implausible high-frequency fluctuations in attention across neighboring spectral positions, we introduce a discrete smoothness penalty along the key dimension:

$$TV_{b,h,i} = \sum_{k=1}^{K-1}\left|\alpha_{b,h,i,j+1} - \alpha_{b,h,i,j}\right| \tag{6}$$

The smoothness loss averages this penalty across batches, attention heads, and query positions and scales it by a regularization weight $\lambda_{\text{smooth}}$:

$$\mathcal{L}_{smooth}(\boldsymbol{\theta}) = \lambda_{smooth}\frac{1}{BHQ}\sum_{b=1}^{B}\sum_{h=1}^{H}\sum_{i=1}^{Q}TV_{b,h,i} \tag{7}$$

With these terms introduced, the final optimization problem is:

$$\boldsymbol{\theta}^{\star} = \arg\min_{\theta}\left[\mathcal{L}_{pred}(\boldsymbol{\theta}) + \mathcal{L}_{\tau}(\boldsymbol{\theta}) + \mathcal{L}_{smooth}(\boldsymbol{\theta})\right] \tag{8}$$

Here, $\mathcal{L}_{\text{pred}}$ ensures accurate regression of the target property, $\mathcal{L}_{\tau}$ regulates the concentration of the attention distribution, and $\mathcal{L}_{\text{smooth}}$ enforces coherence of attention across neighboring positions. Together, these terms constrain the attention mechanism during training, promoting distributions that are both predictive and interpretable.

When the attention regularization terms are enabled, this objective jointly optimizes predictive accuracy and attention interpretability by constraining the concentration and smoothness of the attention distribution. When the regularization terms are disabled, the formulation reduces exactly to standard attention pooling optimized solely through the prediction loss. We refer to this limiting case as *plain attention* throughout the paper.

### II F. Training algorithm

**Table 1**: Shows the algorithm of the proposed method

| |
|---|
| **Require:** dataset $D = \{(x_i, y_i)\}_{i=1}^{N}$, epochs T, batch size m, optimizer, hyperparameters $\lambda_\tau, \lambda_{smooth}$ |
| 1: Initialize network parameters $\theta$ |
| 2: **for** $t = 1$ to $T$ **do** |
| 3: **for** each mini batch $B = \{(x_j, y_j)\}_{j=1}^{m}$ **do** |
| 4: Forward pass: compute predictions $\hat{y_j} = f_\theta(x_j)$ |
| 5: Compute prediction loss $\mathcal{L}_{pred}$ |
| 6: Extract attention weights α from forward pass |
| 7: Compute temperature regularization $\mathcal{L}_\tau$ |
| 8: Compute smoothness loss $\mathcal{L}_{smooth}$ |
| 9: Combine loss: $\mathcal{L} = \mathcal{L}_{pred} + \mathcal{L}_\tau + \mathcal{L}_{smooth}$ |
| 10: Backpropagate $\nabla_{\theta\mathcal{L}}$ and update $\theta$ |
| 11: **end for** |
| 12: Optionally evaluate on validation set and save loss/attention plots |
| 13: **end for** |
| 14: **return** trained parameters $\theta$ |

The complete procedure is summarized in Table 1. The model is trained end-to-end using stochastic gradient-based optimization. During each training iteration, input spectra are tokenized and passed through the encoder to obtain latent feature representations, from which attention logits are computed. The attention distribution is obtained by applying a temperature-scaled softmax, and attention-weighted pooling is used to form a global representation that is passed to the regressor to predict the target property.

The training objective consists of prediction loss together with the attention concentration and smoothness regularization terms. Gradients of the combined loss with respect to all model parameters, including the attention temperature $\tau$, are computed via backpropagation, and parameters are updated using the Adam optimizer[75]. When attention regularization terms are disabled, training reduces exactly to standard attention pooling optimized solely with the prediction loss.

## III. Results

### III A. Synthetic data

To illustrate our approach on a simple synthetic case, we constructed a dataset of 500 spectra composed of three Gaussian peaks with randomly sampled parameters. To mimic experimental conditions, we added white noise to the generated spectra (Figure 2a). The resulting synthetic spectrum:

$$S(x) = \sum_{i=1}^{3} A_i \exp\left(-\frac{(x-\mu_i)^2}{2\sigma_i^2}\right) + \varepsilon \tag{9}$$

Where, $A_i \sim Uniform(0.5, 2.0)$, $\mu_i$ and $\sigma_i$ are sampled from uniform ranges for position and width, and $\varepsilon \sim Normal(0, 0.05)$ is Gaussian noise. The amplitude of the third peak was used as the target variable for the transformer regressor model.

Throughout this section, plain attention refers to the transformer regressor trained using the prediction loss alone, without any explicit regularization on the attention distribution. Regularized attention refers to the same architecture trained under the full SPEAR objective, incorporating both attention concentration control and smoothness regularization. This distinction enables a direct comparison of attention interpretability while holding the predictive model constant.

**Figure 2** compares plain attention and SPEAR models on a synthetic spectral regression task, where the target 'property' is defined as the amplitude of the third Gaussian peak. In **Fig. 2(a)**, global attribution profiles derived from intrinsic attention weights (top) and SHAP values (bottom), included as a post-hoc reference are overlaid on the mean validation spectrum. The plain attention model assigns substantial attention across all three prominent peaks, indicating that attention broadly tracks peak prominence rather than isolating the supervised target. In contrast, SHAP attribution for the plain model is concentrated almost exclusively on the third peak, correctly identifying it as the dominant contributor to the prediction. This discrepancy highlights that, in the unregularized case, attention does not faithfully reflect the true predictive sensitivity of the model.

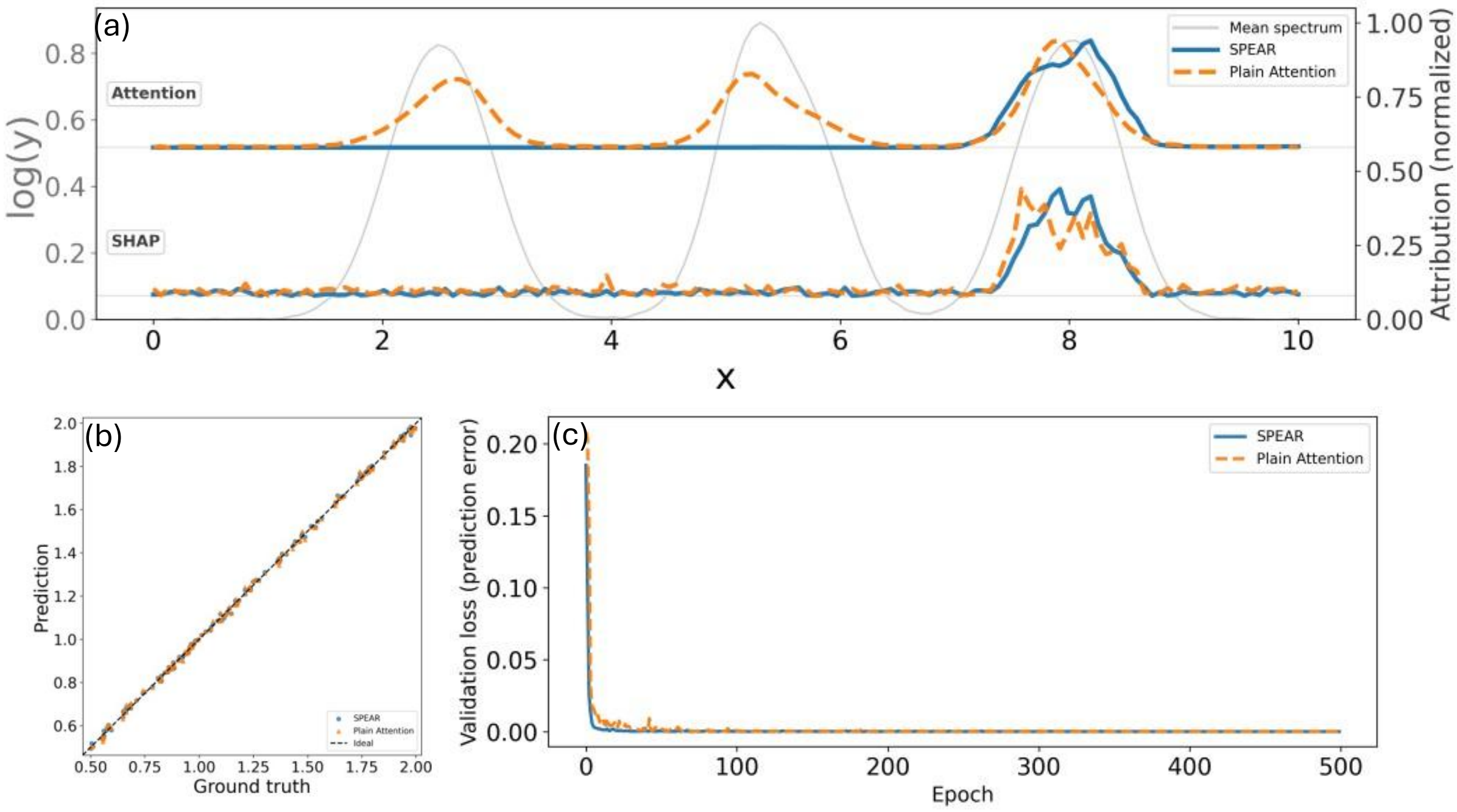

**Figure 2.** Comparison of plain and SPEAR models on a synthetic XRD regression task with a known ground-truth target. **(a)** Mean synthetic spectrum with global attributions from plain (orange, dashed) and SPEAR (blue) models, shown using attention weights (top) and SHAP values (bottom). **(b)** Parity plot comparing predicted and true target values. **(c)** Validation prediction loss versus training epoch.

In contrast, the SPEAR model produces a smooth and contiguous attribution profile that spans the full width of the target peak. The resulting attention distribution closely matches the physically continuous nature of diffraction features, demonstrating that regularization enforces spatial coherence without introducing spurious emphasis on neighboring peaks. The agreement between attention-based attribution and SHAP provides independent evidence that the learned attention distribution reflects the model's predictive behavior rather than an arbitrary visualization artifact.

Importantly, this improvement in interpretability does not come at the cost of predictive performance. As shown in **Fig. 2(b)**, parity plots for both models closely follow the ideal diagonal, indicating accurate recovery of the target peak amplitude across the validation set. **Fig. 2(c)** shows the evolution of validation loss during training, where both models converge rapidly and achieve nearly identical final errors, with the SPEAR model exhibiting slightly smoother and more stable convergence at early epochs.

Together, these results demonstrate that attention regularization reshapes where the model attends without limiting what it can predict. By suppressing high-frequency attribution artifacts and enforcing spatially coherent attention, the SPEAR model yields explanations that are spatially coherent, concentrated on the target peak and consistent with the known generative structure of the synthetic data while retaining full regression accuracy. This separation between explanatory behavior and predictive capacity motivates the parameter sweep study that follows, where we systematically examine how attention smoothness and concentration are controlled by regularization strength.

**Figure 3** illustrates how the strength of attention regularization governs the balance between predictive accuracy and interpretability. **Fig. 3(a)** maps the validation error across a range of regularization parameters, revealing a well-defined intermediate regime in which prediction error is minimized. Weak or absent regularization results in higher error and unstable attribution patterns, while excessive regularization leads to degraded performance.

The corresponding attention distributions for representative cases are shown in **Fig. 3(b)**. In the unregularized model, attention is highly fragmented and exhibits sharp, bin-level fluctuations across the peak, indicative of sensitivity to noise and incidental correlations. Introducing a balanced level of regularization yields a smooth and localized attention profile that spans the physically relevant diffraction peak while suppressing spurious contributions elsewhere in the spectrum. This regime aligns closely with the extended nature of diffraction features and corresponds to the lowest validation error observed in **Fig. 3(a)**.

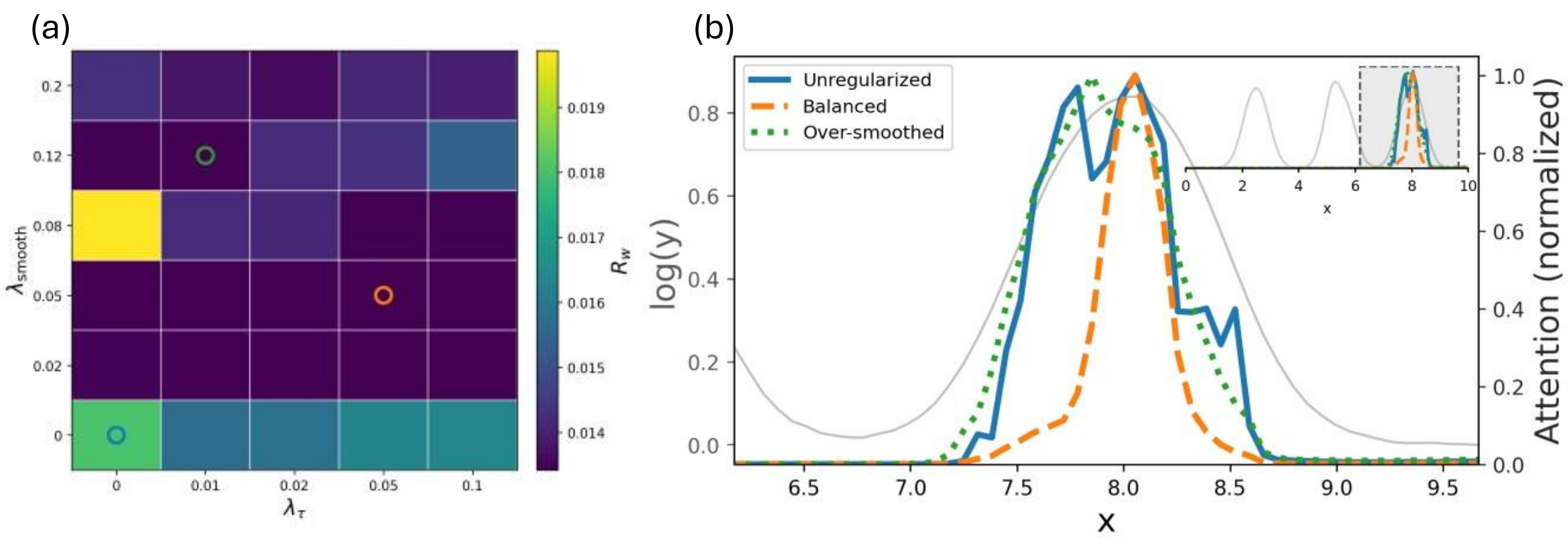

**Figure 3. (a)** Validation error (Rw) as a function of the attention regularization parameters $\lambda_{smooth}$ and $\lambda_\tau$, illustrating the trade-off between under-regularized, balanced, and over-smoothed regimes; highlighted points indicate representative parameter settings discussed in the text. **(b)** Mean XRD spectrum (gray) overlaid with global attention distributions for three representative cases: unregularized attention (blue), balanced regularization (orange), and over-smoothed attention (green, dotted). The inset shows the full spectrum, with the dashed box indicating the region shown in detail.

In contrast, over-smoothing the attention distribution leads to excessively broad attributions that extend beyond the peak of interest. While such attention profiles remain smooth, they dilute selectivity and reduce the model's ability to isolate the structural feature most relevant to the target property, resulting in increased prediction error. Together, these results demonstrate that attention regularization introduces a controllable trade-off: moderate regularization enhances both interpretability and accuracy, whereas insufficient or excessive regularization degrades one or both. This highlights the importance of tuning attention constraints to reflect the physical scale of meaningful features in the data rather than enforcing smoothness indiscriminately.

### III B. Experimental data – Peak Feature

**Figure 4(a)** shows representative diffraction patterns from five wafer locations, with the corresponding positions and target geometry indicated in the inset. The patterns share the same set of reflections across the library, consistent with a single defect-fluorite phase throughout, but differ systematically in peak position and in relative peak intensity with position on the wafer. These variations, rather than the appearance or disappearance of reflections, are the features from which the descriptors below are extracted. To visualize these spatial trends, **Figure. 4(b–e)** maps diffraction-derived descriptors extracted from the spectra across the wafer, including the positions of the first three diffraction peaks ($2\theta$) and the extracted lattice parameter. These maps highlight smooth, reproducible gradients across the library, consistent with underlying compositional variation. Note the center region is ignored due to an experimental artefact that changes the orientation of the center compositions (Ref. 73 for details).

**Figure 5** compares plain attention and SPEAR models applied to experimental XRD spectra, where the target property is the $2\theta$ position of the second diffraction peak. Panel (a) highlights clear qualitative differences in how the two models assign attribution across diffraction features, as assessed using both intrinsic attention weights (top) and SHAP values (bottom). In the plain attention model, attribution is broadly distributed across nearly all prominent diffraction peaks, with assigned importance closely tracking peak intensity. Both attention and SHAP emphasize stronger reflections regardless of their relevance to the target quantity, indicating that the model effectively performs a soft reweighting of the full spectrum rather than isolating the causally relevant feature.

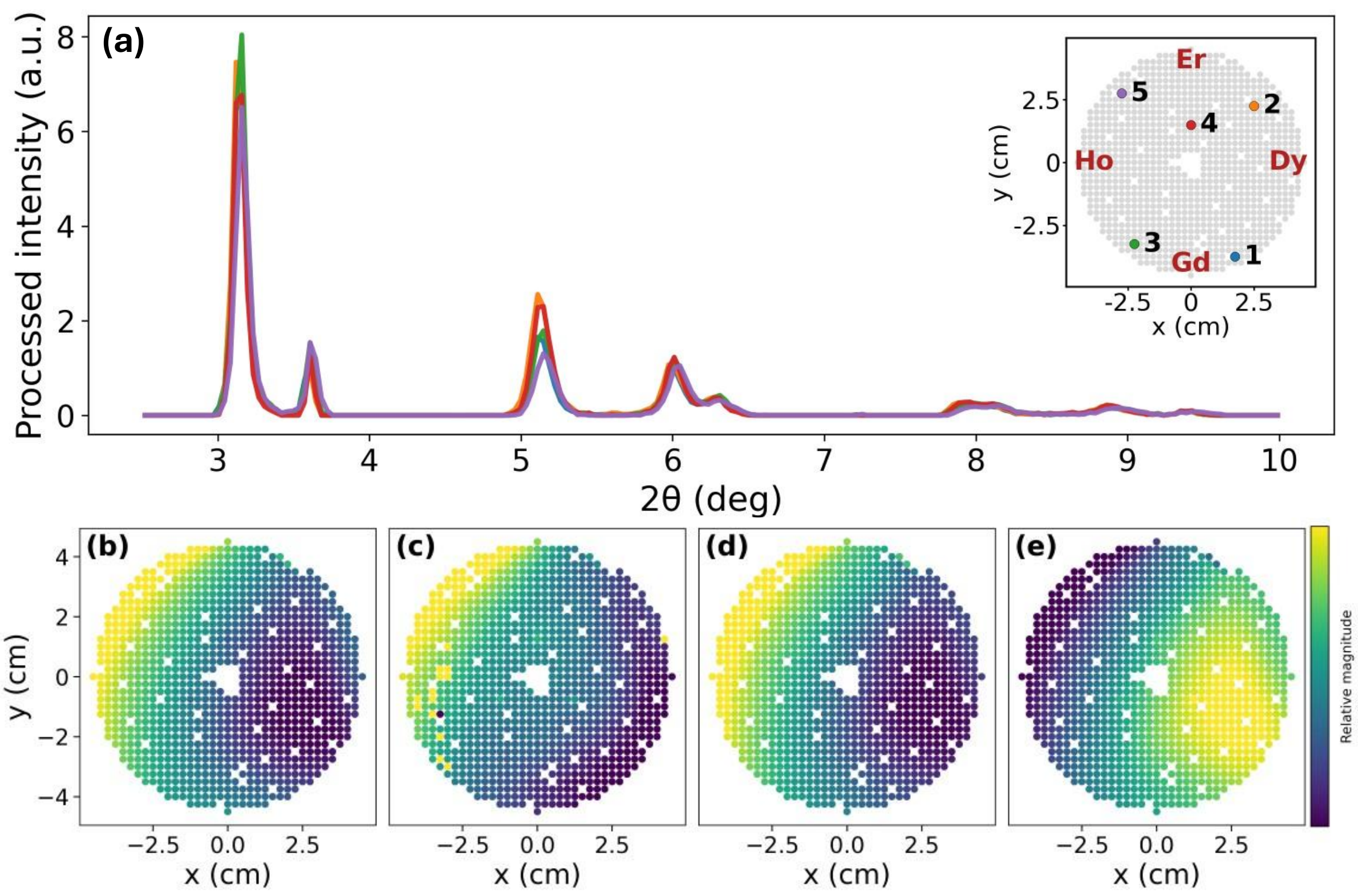

**Figure 4. (a)** XRD spectra from selected wafer locations, with corresponding spatial positions shown in the inset. **(b–e)** Spatial distributions of the peak positions of {111}, {200}, {220} reflections and the determined lattice parameter across the combinatorial wafer. Colors represent relative feature magnitude within each map

In contrast, the SPEAR model exhibits a markedly more selective attribution pattern. Both attention and SHAP concentrate strongly on a single diffraction peak corresponding to the second peak whose position defines the regression target while substantially suppressing attribution to other peaks, even when their intensities are comparable or larger. This behavior demonstrates that the regularized model decouples feature importance from raw peak magnitude and instead concentrates on the spectral region most directly correlated with the target property. The resulting attribution is localized in feature space and smoothly distributed across the width of the relevant peak, consistent with the finite physical extent of diffraction features.

**Figure. 5(b)** and **5(c)** show that this increased selectivity does not come at the expense of predictive performance. Both models achieve accurate predictions, as indicated by parity plots closely following the ideal diagonal. However, the regularized attention model converges more rapidly during training, as evidenced by the faster reduction in validation loss at early epochs. This suggests that constraining attention to emphasize the most informative spectral region simplifies the learning task, enabling more efficient identification of the underlying structure–property relationship..

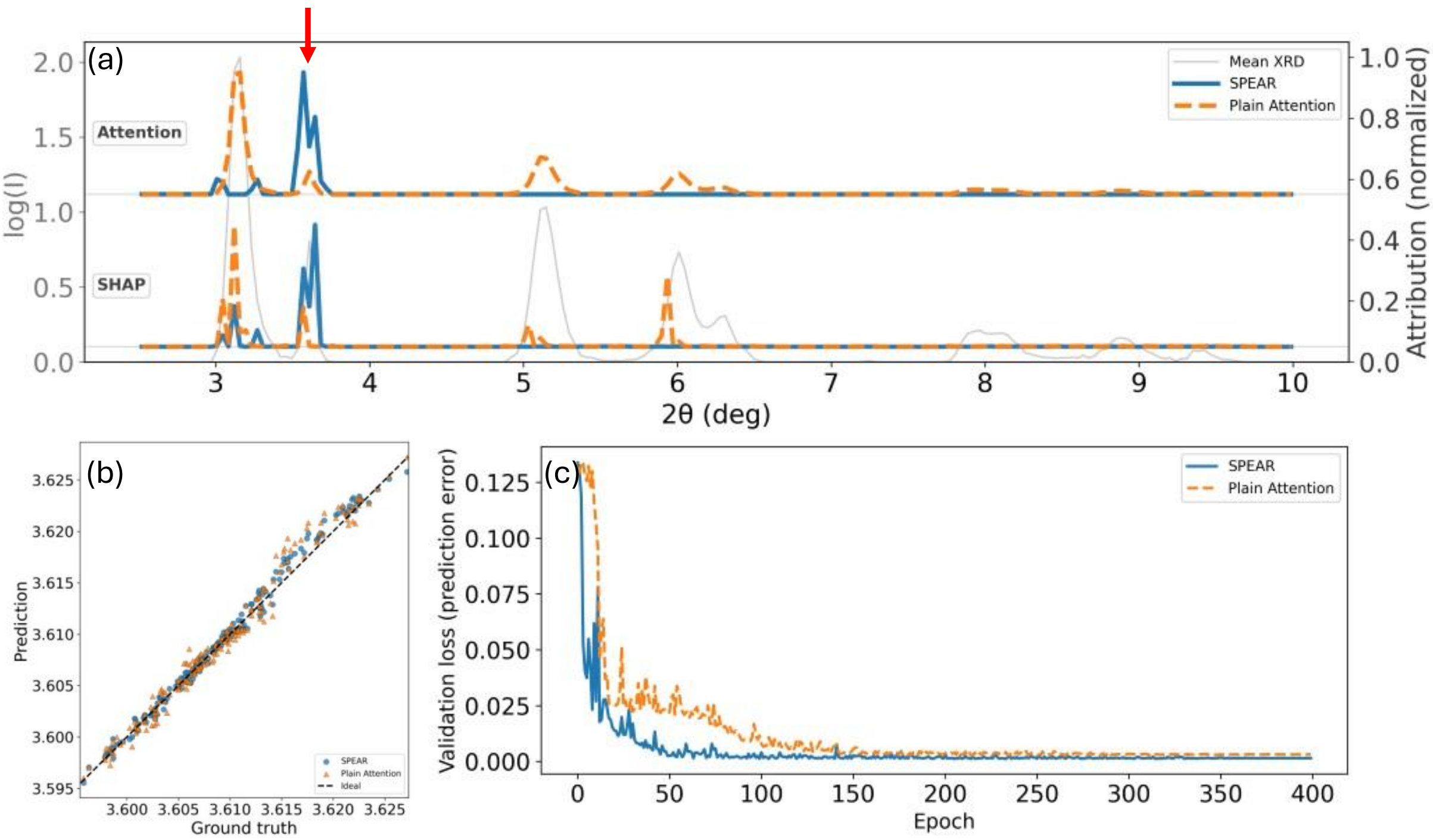

**Figure 5.** Comparison of plain attention and SPEAR models on experimental XRD spectra, where the target property is the $2\theta$ position of the $\{200\}$ diffraction peak **(a)** Mean XRD spectrum with attribution from attention (top) and SHAP (bottom) for plain (orange, dashed) and SPEAR (blue) models. **(b)** Parity plot of predicted versus measured target values. **(c)** Validation loss versus training epoch.

Overall, the behavior of the SPEAR model closely reflects the physical origin of the target quantity. When a material property is explicitly linked to the position of a specific diffraction peak, an interpretable model should preferentially attend to that feature rather than distributing importance across unrelated reflections. Regularized attention achieves this behavior, yielding attention patterns that are spatially localized, physically interpretable, and consistent with the known target, whereas plain attention largely follows peak prominence and produces broader, less informative attributions.

### III C. Experimental data – Material Property

**Figure 6** compares plain attention (unregularized) and SPEAR models applied to experimental XRD spectra for predicting thermal conductivity across the combinatorial material system. In the thermal conductivity task, the target property depends on multiple structural contributions rather than a single peak.

**Figure 6(a)** contrasts attribution patterns obtained from attention weights (top) and SHAP values (bottom). In the plain attention model, attribution remains broadly distributed across most diffraction peaks, with importance largely correlated with peak intensity. Both attention and SHAP assign substantial weight to multiple strong reflections, indicating that the model's attention does not isolate which specific spectral regions are driving the prediction, instead distributing weight broadly in a way that reflects the correlated nature of the diffraction pattern but offers limited guidance for physical interpretation.

In contrast, the SPEAR model exhibits a markedly different attribution structure. Attention and SHAP both concentrate on a limited subset of diffraction features, while other peaks receive reduced weight despite comparable or higher intensities. The agreement between intrinsic attention weights and post-hoc SHAP attribution suggests that the learned attention reflects genuine predictive sensitivity rather than a visualization artifact. The emphasized regions correspond to diffraction planes known to be sensitive to lattice disorder and bonding environments, which are known contributors to phonon scattering and reduced thermal conductivity.

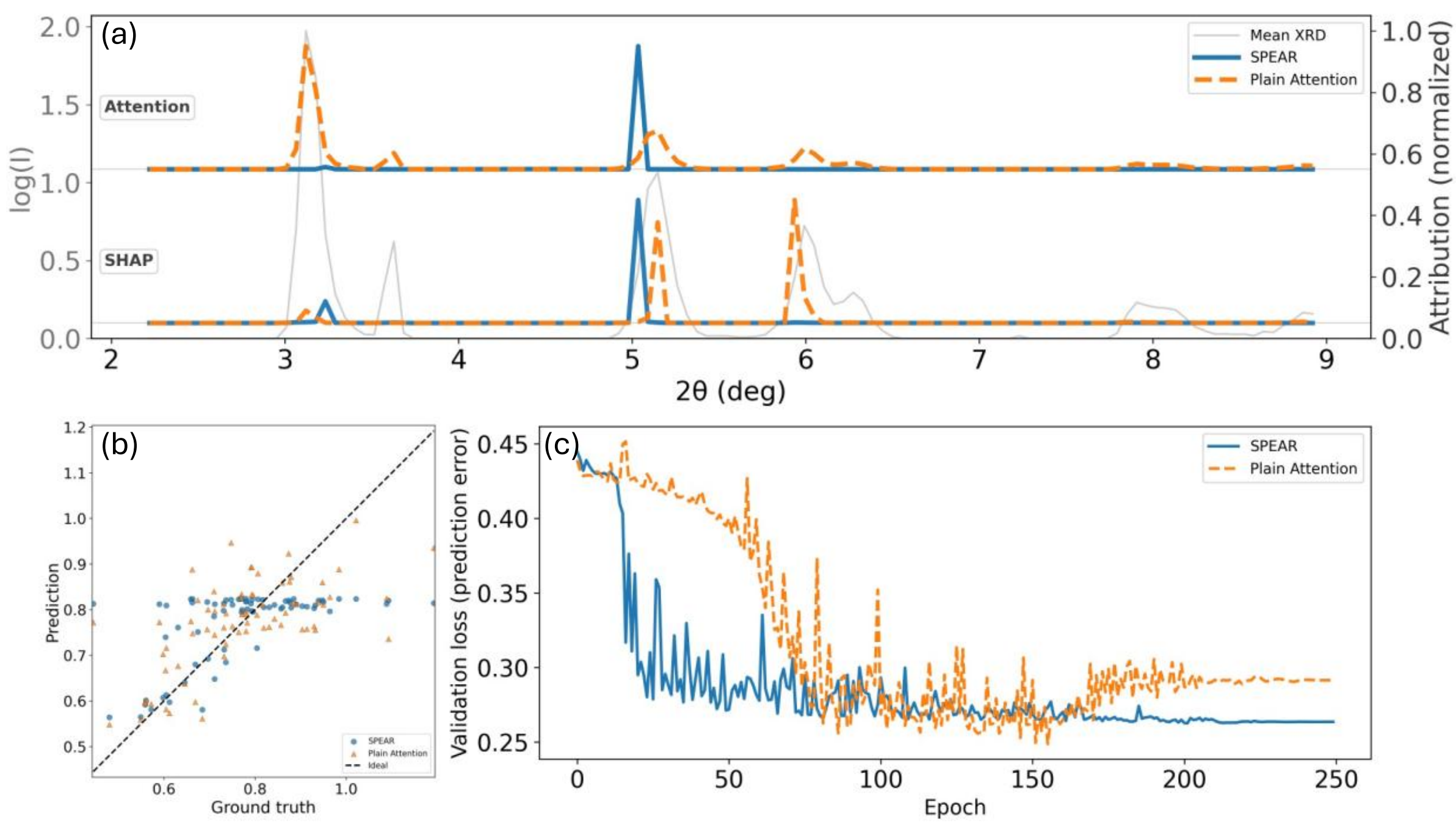

**Figure 6.** Comparison of plain and SPEAR models on experimental XRD spectra, where the target property is the Thermal Conductivity of the corresponding composition. **(a)** Mean XRD spectrum with attribution from attention (top) and SHAP (bottom) for plain (orange, dashed) and SPEAR (blue) models. **(b)** Parity plot of predicted versus measured target values. **(c)** Validation loss versus training epoch.

**Figure 6(b)** shows predicted versus ground-truth thermal conductivity for both models. While both approaches achieve accurate predictions, the SPEAR model exhibits reduced scatter, indicating improved stability across the validation set. **Figure 6(c)** further demonstrates that the regularized attention model converges more rapidly during training and achieves a slightly lower validation loss, consistent with the model focusing on a smaller, more informative set of spectral regions.

Together, **Figure 6** demonstrates that regularized attention not only improves interpretability in controlled settings (Figures 2 and 5) but also yields consistent, physically meaningful attribution in complex, real-world structure–property prediction tasks where multiple spectral features jointly determine the target quantity.

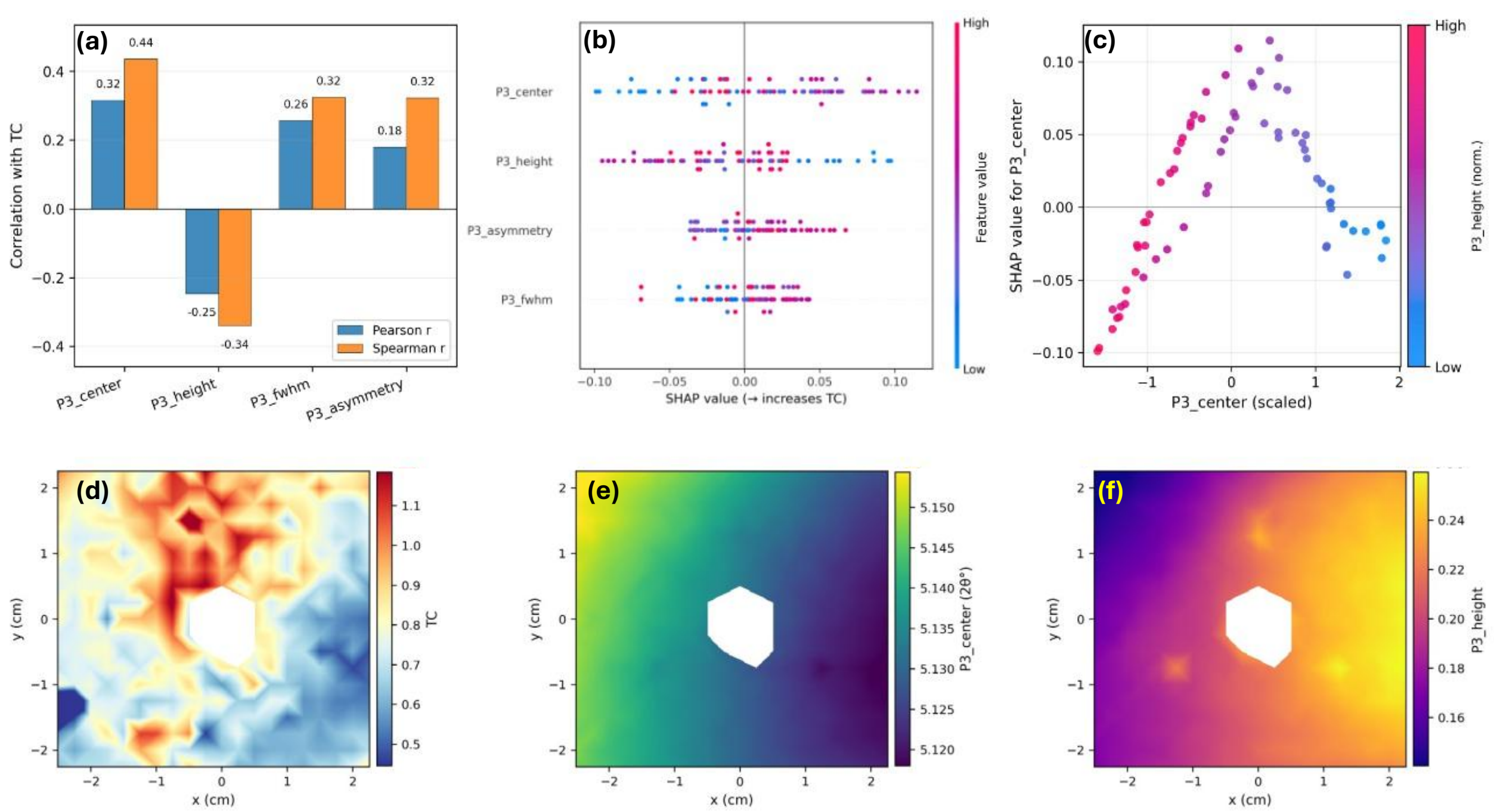

**Figure 7.** Structural drivers of thermal conductivity at the SPEAR-identified reflection. **(a)** Pearson and Spearman correlation coefficients between pseudo-Voigt peak descriptors and TC. **(b)** SHAP beeswarm plot showing feature importance and direction **(c)** SHAP dependence plot for peak center position, colored by peak amplitude **(d-f)** Spatial maps of TC, peak center position, and peak amplitude across the combinatorial wafer.

**Figure 7** examines which properties of the SPEAR-identified reflection, the {220} (denoted P3 in the panel labels), carry the thermal conductivity signal. Panel (a) shows that the peak center position is the most strongly correlated descriptor (Pearson r = 0.32, Spearman ρ = 0.44), followed by FWHM and asymmetry, while peak height is the only descriptor anti-correlated with TC (r = −0.25, ρ = −0.34). Spearman coefficients exceed Pearson values for every descriptor, indicating monotonic but nonlinear relationships. The SHAP beeswarm in panel (b) reproduces this ordering and confirms the direction of each effect. Panel (c) resolves the nonlinearity: the SHAP contribution of the peak position is itself non-monotonic in the peak position, rising to a maximum at intermediate positions and declining at both extremes of the observed range. The coloring by peak amplitude shows that high amplitudes coincide with low peak positions and vice versa, confirming that the two descriptors vary inversely across the library. The spatial maps in (d–f) place these descriptors on the wafer, where the peak position and amplitude gradients run counter to one another and the thermal conductivity map retains additional structure not captured by either descriptor alone.

These observations show that the relationship between P3 and TC is complex, and connects directly to the structural picture developed in our earlier synchrotron study.[73] There, the tetragonal distortion identified from the pseudocubic {200}/{002} reflections (peak 2) was minimized in the low thermal conductivity region. Concomitantly, while the largest lattice parameter was expected to be nearest the Gd target location, the synchrotron diffraction revealed the largest lattice parameter in the Gd/Dy quadrant, consistent with the local thermal conductivity minima. Together this suggested an interplay in the size disorder accommodation that has interesting effects on the thermal conductivity. The SPEAR results add a new layer to our analysis, as we previously missed that the 220 peak position, together with its width and asymmetry, varies systematically across the library in a manner consistent with the tetragonal distortion identified in our earlier synchrotron analysis, and thus with thermal conductivity. Several factors may explain why the attention mechanism preferentially emphasized the {220} reflection over the more commonly examined {200} family. The higher diffraction angle increases sensitivity to small lattice distortions through larger angular shifts, while the progressive broadening and asymmetry of the {220} peak provide a continuous signature of symmetry lowering. Furthermore, if the surrounding spectral region contains fewer overlapping features or a lower background, the resulting signal-to-noise ratio may further enhance its discriminative value. These factors collectively suggest that the model has identified a region of high structural information content rather than simply reproducing conventional crystallographic heuristics.

From a film growth perspective, orientation selection in the sister material yttria stabilized zirconia (YSZ) has been studied extensively. Sonnenberg et al.[76] found that ion bombardment during growth induces alignment through an orientation-dependent growth instability rather than through preferential etching of misoriented nuclei. Additionally, Mahieu et al.[77] subsequently related in-plane alignment in magnetron sputtered YSZ to the angular spread of the incoming material flux, and in a study of the $Ar/O_2$ ratio[78] showed that increasing the oxygen partial pressure shifts the out-of-plane texture from nearly pure {002} toward a mixture of approximately 70% {111} and 30% {220}. We note that substrate bias does change the orientation of our films from the 200 preferred orientation to a pattern closer to one expected from a randomly oriented film. Preferred orientation acts on relative intensities rather than directly on peak positions, and the intensity behavior of this reflection is captured in the peak height descriptor. The variation in the {220} peak shape across the library therefore cannot be explained by changes in peak height alone, although a contribution from texture cannot be excluded: where the tetragonal components are unresolved, texture can alter their relative weights and thereby shift the fitted center, width, and asymmetry. A quantitative assessment would require normalized integrated intensities or texture coefficients rather than fitted peak heights. We suggest that the observed behavior reflects the tetragonal distortion that accommodates cation size disorder in these defect fluorite films, a hypothesis that texture-resolved or pole-figure measurements across the composition gradient would be well suited to test.

Overall, these results demonstrate that introducing learnable regularization into the attention mechanism enhances interpretability without sacrificing predictive performance, and that the resulting attribution can generate a physically grounded, testable hypothesis about the structural origin of the target property rather than simply highlighting the most prominent spectral features.

**Conclusion**

In this work, we investigated attention-based regression models for structure–property learning in materials systems, with a focus on improving the interpretability and stability of attention weights without compromising predictive performance. While attention mechanisms are often treated as inherently interpretable, we show that unregularized attention can yield fragmented, intensity-driven, and unstable attention patterns that are difficult to interpret in terms of underlying physics, particularly in spectroscopic and diffraction data.

To address this limitation, we introduced SPEAR, a learnable attention framework incorporating explicit regularization terms that control both the concentration and smoothness of attention distributions. Using synthetic benchmarks with known generative relationships, we demonstrated that attention regularization reshapes attribution patterns into smooth, contiguous profiles aligned with the underlying physical features, while maintaining equivalent regression accuracy. Systematic ablation studies revealed a well-defined intermediate regime in which attention regularization improves both interpretability and predictive stability, whereas insufficient or excessive regularization degrades attribution quality and/or performance.

We further applied the approach to experimental XRD data from a combinatorial $(\mathrm{GdDyHoEr})_2\mathrm{Zr}_2\mathrm{O}_7$ thin-film library. In this setting, the SPEAR model exhibited selective, target-aware attribution, preferentially focusing on the diffraction peak directly associated with the target property rather than broadly weighing all high-intensity features. This behavior contrasts with plain attention, which primarily tracks peak prominence, and highlights the ability of regularized attention to decouple physical relevance from raw signal intensity. Importantly, the improved attribution structure was accompanied by comparable or modestly improved convergence behavior and prediction error. Beyond validating the method, the attribution proved scientifically generative: the model's emphasis on the {220} reflection prompted us to re-examine our earlier synchrotron analysis, revealing a correlation between the {220} peak position, the tetragonal distortion, and the local thermal conductivity that we had not previously identified.

Overall, these results establish attention regularization as a physically motivated inductive bias that improves the interpretability of attention-based models for structure–property regression. Rather than treating attention as a post hoc visualization tool, the proposed framework embeds interpretability constraints directly into the learning process, yielding explanations that are more stable, selective, and physically interpretable for diffraction data from disorder-dominated materials systems. This approach is broadly applicable to other spectroscopic, imaging, and spatially structured datasets and provides a foundation for using attention as a reliable guide for hypothesis generation in high-throughput and data-driven materials discovery.

## Acknowledgments

This research is supported (U.P, A.R, D.A.P, J.H, K.P, P.D.R and S.V.K) by the National Science Foundation Materials Research Science and Engineering Center program through the UT Knoxville Center for Advanced Materials and Manufacturing (DMR-2309083). This research used resources (28-ID-1) of the National Synchrotron Light Source II, a U.S. Department of Energy (DOE) Office of Science User Facility operated for the DOE Office of Science by Brookhaven National Laboratory under Contract No. DE-SC0012704.

## Data and code availability

All data and code supporting this study are available at https://github.com/adityaraghavan98/SPEAR.